\documentclass[12pt,preprint]{aastex}

\usepackage{graphicx}% Include figure files
\DeclareGraphicsExtensions{.pdf, .jpg, .png}

\newcommand{\FITDisk}{{\sc FITDisk}}
\newcommand{\br}{{\bf r}}
\newcommand{\bvv}{{\bf v}}
\newcommand{\Porb}{P_{\rm orb}}
\def\mean#1{{\langle}#1{\rangle}}

\begin{document}
\singlespace

%\preprint{APS/123-QED}

\title{FITDisk: A Cataclysmic Variable Accretion Disk Demonstration Tool}

\author{Matt A. Wood, Josh Dolence}
\affil{%
Department of Physics \& Space Sciences and The SARA Observatory\\
Florida Institute of Technology, Melbourne, FL 32901}
 \email{wood@fit.edu; jdolence@fit.edu}
 
\author{James C. Simpson}
\affil{%
NASA YA-D7, Kennedy Space Center, FL 32899}
\email{James.C.Simpson@nasa.gov}

\slugcomment{Submitted to PASP 2005 Nov 7; 
Accepted 2005 Nov 11}% It is always \today, today,

\begin{abstract}
We present the software tool {\FITDisk}, a precompiled-binary Windows GUI version of our smoothed particle hydrodynamics cataclysmic variable accretion disk research code.  Cataclysmic variables are binary star systems in which a compact stellar remnant, typically a white dwarf star, is stripping mass from a lower-main-sequence companion star by way of an accretion disk.  Typically the disk is the brightest component of the system, because the plasma is heated dramatically as it spirals down in the gravitational well of the primary white dwarf star.  The shortest-period systems can display disk ``superhump'' oscillations driven by the rotating tidal field of the secondary star.  \FITDisk\ models these accretion disk phenomena using a fully three-dimensional hydrodynamics calculation, and data can be visualized as they are computed or stored to hard drive for later playback at a fast frame rate.  Simulations are visualized using OpenGL graphics and the viewing angle can be changed interactively.  Pseudo light curves of simulated systems can be plotted along with the associated Fourier amplitude spectrum.  \FITDisk\ is available for free download at {\tt www.astro.fit.edu/cv/fitdisk.html}.
\end{abstract}

\keywords{accretion disks
-- binaries: close 
-- cataclysmic variables
-- methods: numerical
-- hydrodynamics
}

\section{An Introduction to Cataclysmic Variables}

Cataclysmic variable stars are very close binary star systems consisting of a compact stellar remnant {\it primary} star -- typically a white dwarf star -- which is stripping mass from a less-massive lower-main-sequence {\it secondary} star (Hellier 2001, Warner 1995).  They are closely related to the low-mass X-ray binaries, which instead have neutron star or black hole primaries.  Cataclysmic variables display a wide range of behavior, the most dramatic being the {\it classical nova} explosion which occurs when the accreted surface hydrogen layer is massive enough to ignite a thermonuclear runaway in the envelope layers of the primary white dwarf star, causing the system to brighten by a factor of $\sim$$10^4$ to $10^6$ over the pre-eruption brightness.  Less dramatic but far more common are the photometric variations caused by the accretion disk itself, which is typically the brightest component of the system. Simple energy arguments (Frank, King, \& Raine 1992) show that the total accretion disk luminosity is approximately
\begin{equation}
L_{\rm acc} \approx {G M_1 \dot M \over R_1},
\end{equation}
where $M_1$ and $R_1$ are the white dwarf mass and radius, respectively, $G$ is the gravitational constant, and $\dot M$ is the accretion rate onto $M_1$.  Roughly speaking, half the luminosity should be radiated in the disk, and half in the boundary layer where the Keplerian accretion flow settles onto the more-slowly rotating white dwarf.  Viscosity within the differentially-rotating fluid of the disk acts to transport angular momentum outward in radius, so that mass can migrate inward.  Angular momentum from the outer disk is fed back into the orbital angular momentum of the secondary star via tidal torques.  The material in the accretion stream is highly supersonic when it impacts the edge of the disk, causing a {\it bright spot} which when observed in high inclination (i.e., more nearly edge-on) systems leads to an observed brightening for that portion of the orbit that the bright spot is facing the observer.  This brightening once per orbit is called an {\it orbital hump}, and if present reveals the orbital period.

Systems with high mass transfer rates $\dot M\gtrsim 10^{-9}\ M_\odot\rm\ yr^{-1}$, where $M_\odot$ indicates units of solar masses, are classified as old-novae or novalikes, depending on if a classical nova explosion has been observed or not.  The disks in these systems are thought to be in a permanent high-viscosity state, such that mass transfer through the disk and onto $M_1$ is efficient and in a state of quasi-equilibrium.  Most systems with smaller mass transfer rates $\dot M\lesssim10^{-10}\ M_\odot\rm\ yr^{-1}$ are so-called dwarf nova systems, and have disks that cycle between high and low states.  The mass flow through $L_1$ is not sufficient to keep the disk permanently in high state, so matter accumulates in the disk until a critical point is reached.  Before this critical point is reached, the mass flow rate onto $M_1$ is low, and hence so is the system luminosity.  Once reached, however, a heating wave propagates through the disk material, sharply increasing the viscosity and mass flow through the disk and onto the white dwarf, resulting in a rapid brightening of the system by a factor of $\sim$$10^2$ in what is called a {\it dwarf nova outburst}.  The source of accretion disk viscosity is thought to be the magnetorotational instability (Balbus \& Hawley 1991).

In addition to normal dwarf nova outbursts, many systems with short orbital periods ($\Porb \lesssim 3.1$ h) are observed to display {\it superoutbursts}, which are roughly twice as bright and last some 5 times longer (Patterson et al. 2005) than normal dwarf nova outbursts.  In all superoutbursting systems, photometric oscillations with a period a few percent longer than the orbital period are observed to grow to detectability on a timescale of a few tens of orbits.  Because they are associated with superoutbursts, these periodic signals are known as {\it superhumps}.  We now understand that superhumps are the result of a driven oscillation of the disk by the rotating tidal field of the secondary star (Whitehurst 1988; Hirose \& Osaki 1990; Lubow 1991; Simpson \& Wood 1998).  For systems with mass ratios $q\equiv M_2/M_1 \lesssim 0.34$, the disk can extend out to the region of the 3:1 co-rotation radius, and the inner Lindblad resonance can be excited (Lubow 1991).

Cataclysmic variables have long been a favorite target of amateur and professional astronomers and anyone with a modest-aperture telescope and a CCD camera can obtain useful data on brighter objects.  For example, the {\it Center for Backyard Astrophysics} is a global network of (mostly amature) astronomers with small telescopes dedicated to the study of cataclysmic variables ({\tt cba.phys.columbia.edu}), and the {\it American Association of Variable Star Observers} (AAVSO), founded in 1911 at Harvard College Observatory to coordinate variable star observations made largely by amateur astronomers, has a large contingent interested in CVs specifically ({\tt aavso.org}).

Our group has been studying the superhump phenomenon numerically using the method of smoothed particle hydrodynamics (SPH).  The experiments have confirmed the physical origin of the observed superhump lightcurves (Simpson \& Wood 1998, Wood, Montgomery, \& Simpson 2000), and shown that the light curves should become harmonically more complex for disks observed more nearly edge-on (Simpson, Wood, \& Burke 1998).  Our research code computes the hydrodynamics of accretion disks in three dimensions, and is highly optimized for serial computers, running as a command-line Fortran program under the Linux operating system.  Because of the high level of interest among observers in CVs and the difficulty in visualizing accretion disk dynamics from artist conceptions and text descriptions in journals, we decided to develop and release a demonstration version of \FITDisk\ that includes a graphical user interface (GUI) and which runs under the WindowsXP operating system.

In section 2, we briefly introduce the method of SPH as applied to cataclysmic variable accretion disks.  In section 3 we discuss the porting of  \FITDisk\ to the WindowsXP environment and how to use the program.  We show how our simulation light curves compare with time-series observations of the helium dwarf nova CR Boo in section 4.  Conclusions are given in section 5.

\section{Smoothed Particle Hydrodynamics}

The method of smoothed particle hydrodynamics (SPH) approximates continuum hydrodynamics using a scattered grid, where the grid points are effectively Lagrangian point markers in the fluid (Lucy 1977, Monaghan 1992, 2005).  It is a powerful technique for simulating physical systems that are highly dynamic and bounded by vacuum; thus, for many astrophysical systems, SPH can be a more computationally efficient approach than more conventional Eulerian techniques.  SPH replaces the fluid continuum with a finite number of particles which interact pairwise with each other, where the strength of that interaction is a function of the interparticle distance.  The function is the {\it kernel} $W$, and many choices for the kernel function are possible.  The simplest choice would be a Gaussian, but the consequence of this choice is the interaction force is non-zero for all other particles in the system.  It is common to use a polynomial spline function which approximates a Gaussian, but which has compact support, being identically zero beyond twice the {\it smoothing length} $h$ (Monaghan \& Lattanzio 1985).
With this choice, interparticle forces need only be calculated for neighbors within $2h$ of a given particle.  In the continuum limit, a field quantity can be estimated by the integral
    \begin{equation}
\mean{A({\br})}={\int{W({\br}-{\br^\prime},h)
    {A(\br^\prime)} d{\br^\prime}}},    \label{eq: theo_ave}
    \end{equation}
where the integration is over all space and where $W$ is defined such that $\int W\, dV\equiv1.$  To compute a field quantity using the finite number of SPH particles within $2h$ of $\br$, we have
    \begin{equation}
A({\br})=\sum_j m_j{{A_j}\over{\rho_j}}W({\br-\br_j},h),
    \label{eq: sph_ave}
    \end{equation}
where $m_j$ is the mass of the $j$-th particle, $\rho_j$ is it's density, and $\br_j$ is its position.  A general form for the estimate of the derivative of a field quantity is given by
    \begin{equation}
\nabla A({\br})=\sum_j m_j{{A_j}\over{\rho_j}}\nabla W
    ({\br-\br_j},h).\label{eq: sph_deriv}
    \end{equation}

The general form of the momentum and energy equations relevant for accretion disk studies are
    \begin{eqnarray}
    {d^2{\bf r}\over{dt^2}}&=&
        -{\nabla P\over \rho}
        +{{\bf f}_{visc}}
        -{{GM_1\over {{r_1}^3}}{{\bf r}_1}}
        -{{GM_2\over {{r_2}^3}}{{\bf r}_2}},\\
        {}\nonumber\\
    {{du}\over{dt}}&=&-{P\over\rho}{\nabla\cdot\bf{v}}+{\epsilon_{visc}},
    \end{eqnarray}
where
$u$ is the specific internal energy,
$\bf {f}_{visc}$ is the viscous force,
$\epsilon_{visc}$ is the energy generation from viscous dissipation, and
$\br_{1,2}\equiv \br - \br_{M1,M2}$ are the displacements from stellar masses $M_1$ and $M_2$, respectively.  We assume an ideal gamma-law equation of state, $P=(\gamma-1)\rho u$, where we typically assume $\gamma=1.01$.

Our SPH form of the momentum equation for a given particle $i$ is
    \begin{equation}
    {d^2{\bf r}_i\over{dt^2}}
    = -\sum_j m_j\left({P_i\over{{\rho_i}^2}}
        +{P_j\over{{\rho_j}^2}}\right)\left(1+\Pi_{ij}\right){\nabla_i{W_{ij}}}
        -{{GM_1\over {{r_{i1}}^3}}{{\bf r}_{i1}}}
        -{{GM_2\over {{r_{i2}}^3}}{{\bf r}_{i2}}}.
    \label{sphforce}\end{equation}
where we use a standard prescription for the artificial viscosity (Lattanzio 1986)
\begin{equation}
\Pi_{ij} = \left\{ \begin{array}{ll}
        -\alpha\mu_{ij} + \beta\mu_{ij}^2 & {\bf v}_{ij}\cdot {\bf r}_{ij} \le 0;\\
        0  & \mbox{otherwise;}
        \end{array}\right.
\end{equation}
where
\begin{equation}
\mu_{ij} = {h{\bf v}_{ij}\cdot {\bf r}_{ij} \over c_{s,ij}(r_{ij}^2 + \eta^2)}.
\end{equation}
Here $c_{s,ij}$ is the average of the soundspeeds for particles $i$ and $j$, and we use the notation $\bvv_{ij} = \bvv_i - \bvv_j$, $\br_{ij} = \br_i - \br_j$.
In \FITDisk, we fix $\eta=0.1h$, but the user can select the viscosity coefficients using a slider.  Note that as is typical in artificial viscosity formalisms, only approaching particles are subject to a viscous force.

The internal energy of each particle is integrated using an action-reaction principle (Simpson \& Wood 1998)
    \begin{equation}
{{du_i}\over{dt}}=-{\bf a}_{i,SPH}\cdot{\bf v}_i,\label{eq: ourdudt.1}
    \end{equation}
which is formally equivalent to the standard SPH
internal energy equation but computationally much more efficient.  If this method fails, however, we then use the more standard form
    \begin{equation}{{du_i}\over{dt}}={P_i\over{{\rho_i}^2}}
{{\sum_j}m_j\left(1+\Pi_{ij}\right){{\bf
v}_{ij}}\cdot{\nabla_i{W_{ij}}}}.\label{sphenergy}
    \end{equation}
Like all hydrodynamics calculations, the timesteps in \FITDisk\ are sound-speed limited.  Because physically the orbital period at the surface of a white dwarf is $\sim$10 s, while the orbital period of the outer disk is $\sim$1 h, it is advantageous to let individual particles have timesteps as short as needed.  Timesteps can be as long as $\delta t_0=\Porb/200$, but a particle can have a shorter timestep $\delta t_i= \delta t_0/2^m$ as required by the local environment, where $0\le m\le8$ provides sufficient dynamic range for all simulations.

The most common observation of superhumping cataclysmic variables is time series photometry -- i.e., sampling the brightness hundreds or more times per orbit.  In \FITDisk\
we assume the sum over all particles of the changes in the internal energies over a time step $\delta t_0^n$ is directly proportional to the luminosity of the disk over that same time interval. Thus we can calculate an approximate ``light curve'' for the simulation:
\begin{equation}
L^n = \sum_j \left({du\over dt}\right)_j^n \delta t_0
\end{equation}

\FITDisk\ simulates the fluid dynamics of an accretion disk subject to the gravitational potential of the two stars in the system.  The calculations are made in the inertial frame.  We treat the stars as point masses on circular orbits, and the center of mass of the system is the origin of the coordinate system.  Particles are injected with thermal velocities $T \sim 4000$~K at random coordinates within a small box at the $L_1$ point.  Particles are considered to be lost from the system if they are accreted onto $M_1$ or $M_2$, or are ejected from the system.

Our code specifically models accretion disks in the novalike systems.  Particles are injected in a burst at a rate of a few-thousand per orbit until the desired number of particles is reached.  Whenever a particle is lost, a replacement particle is injected at $L_1$.  Thus, depending on the parameters of the simulation, the disk will either evolve to a state of quasi-equilibrium, or may eventually be driven into superhump oscillations.

\section{USING The GUI FITDISK}

 To port our research code from the Linux operating system to the WindowsXP operating system, we used Compaq Visual Fortran and added a graphical user interface and visualization using OpenGL.  To ensure that the machine requirements (RAM and hard drive space in particular) are as minimal as possible while still allowing interesting results, we limit the total number of particles to 25,000, although superhumps can be observed with as few as 1,000 particles in a run time of under 5 minutes, allowing direct usage in a classroom setting.  The program uses about 100MB of RAM, so we recommend that systems have least 256 MB of RAM for best results.  Recording 100 orbits of a 25,000-particle simulation at high resolution will require approximately 7.5 GB of disk space.

We maintain a webpage for \FITDisk\ at {\tt www.astro.fit.edu/cv/fitdisk.html}, and the executable code and a reference manual can be downloaded from that site.  The download file is a 'zip' archive containing {\tt fitdisk.exe} and two auxillary files in a subfolder.  These will extract to a directory named FITDisk.  The initial release version of \FITDisk\ is v1.0.  We have set up a mechanism for users to send feedback at our website, and will release updates and bug fixes as needed, based on this feedback.

\begin{figure}
\plotone{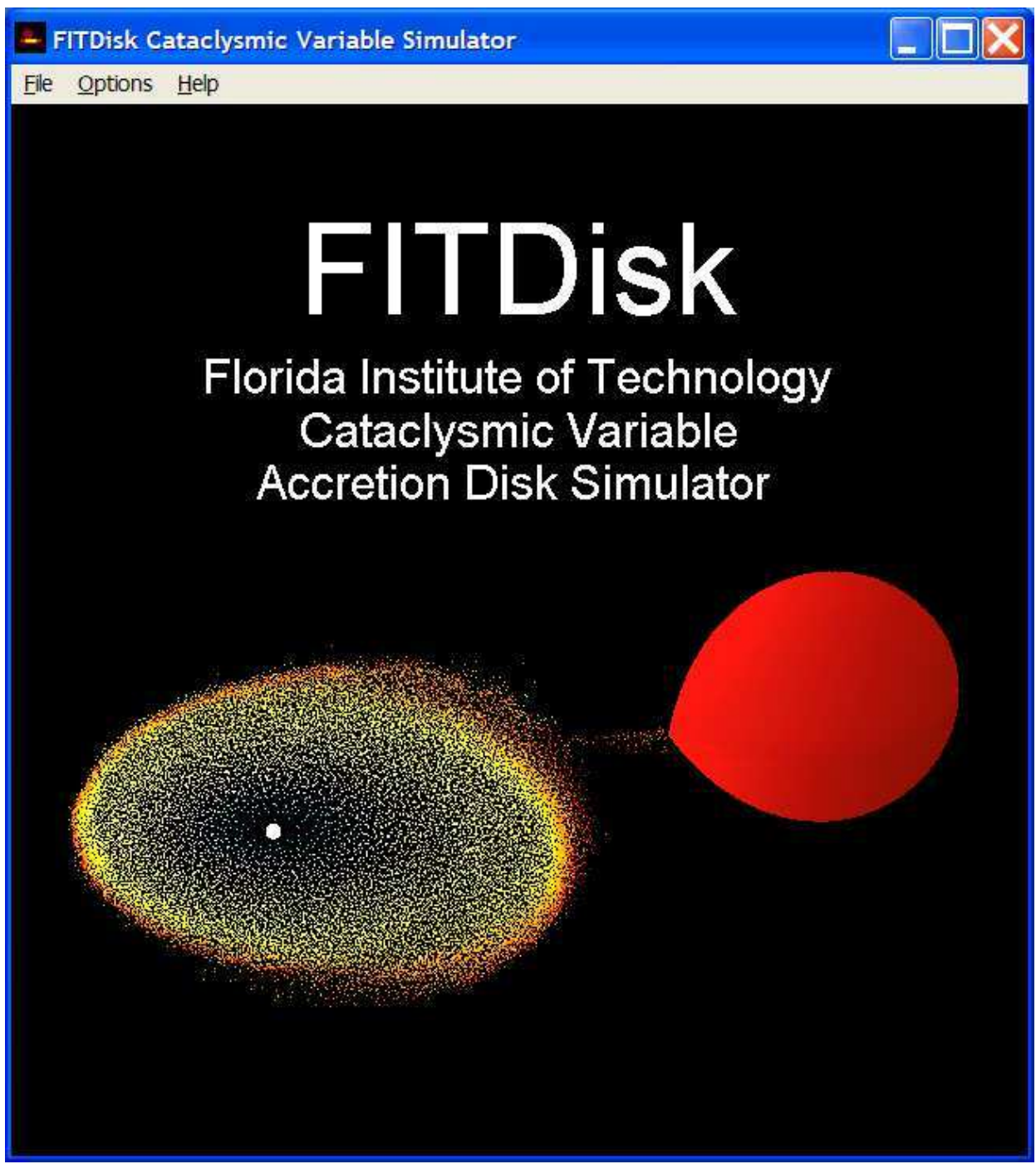}% Here is how to import EPS art
%\includegraphics[scale=0.7]{wood-fig1.png}
%\vspace{6in}
\caption{\label{fig:1st-screen} The opening screen of \FITDisk.  The image shown is that of a 25,000-particle superhump simulation calculated with \FITDisk. The figure may appear in color online.}
\end{figure}

\begin{figure}
\plotone{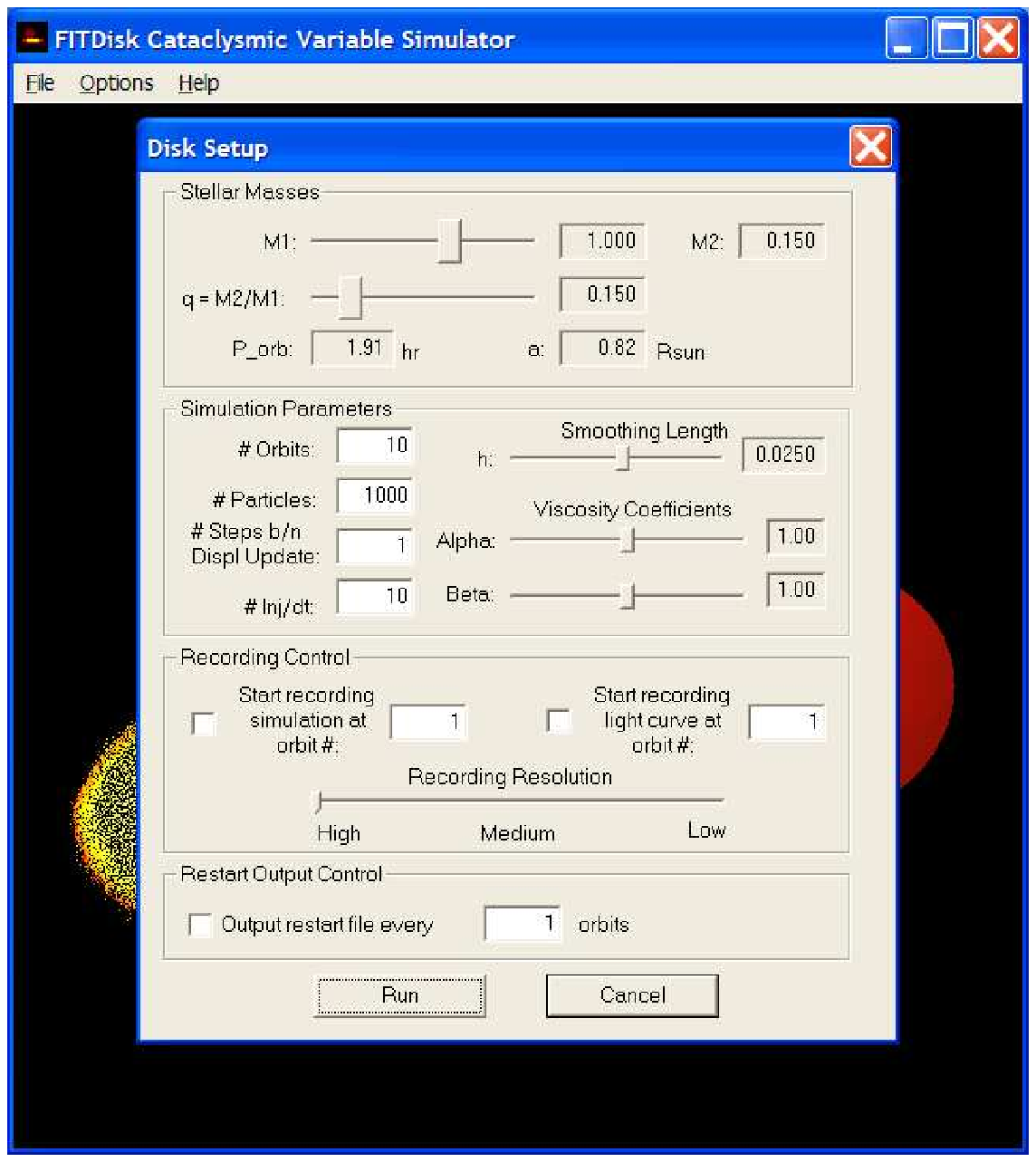}% Here is how to import EPS art
\caption{\label{fig:gui} The Graphical User Interface for starting a new disk.  The default parameters shown will lead to superhump oscillations within 10 orbits.}
\end{figure}

Figure~\ref{fig:1st-screen} shows the startup screen, including a still image from a 25,000-particle simulation. Figure~\ref{fig:gui} shows the GUI that comes up if File$\rightarrow$New Disk is selected.    Table~\ref{tab:table1} lists the adjustable parameters, their ranges, and default values.   \FITDisk\ can be used to explore a significant volume of parameter space. When simulating a relatively small number of particles ($\sim$1000), the code runs fast enough to watch the disk build and evolve to superhumps in real time. However, when calculating a simulation with the maximum number of particles evolved over enough time for superhumps to develop, even users with the latest hardware will probably want to record the results to the hard drive for next-day playback at a fast frame rate.  This is accomplished using standard Windows file requestors.

\begin{figure}
\plotone{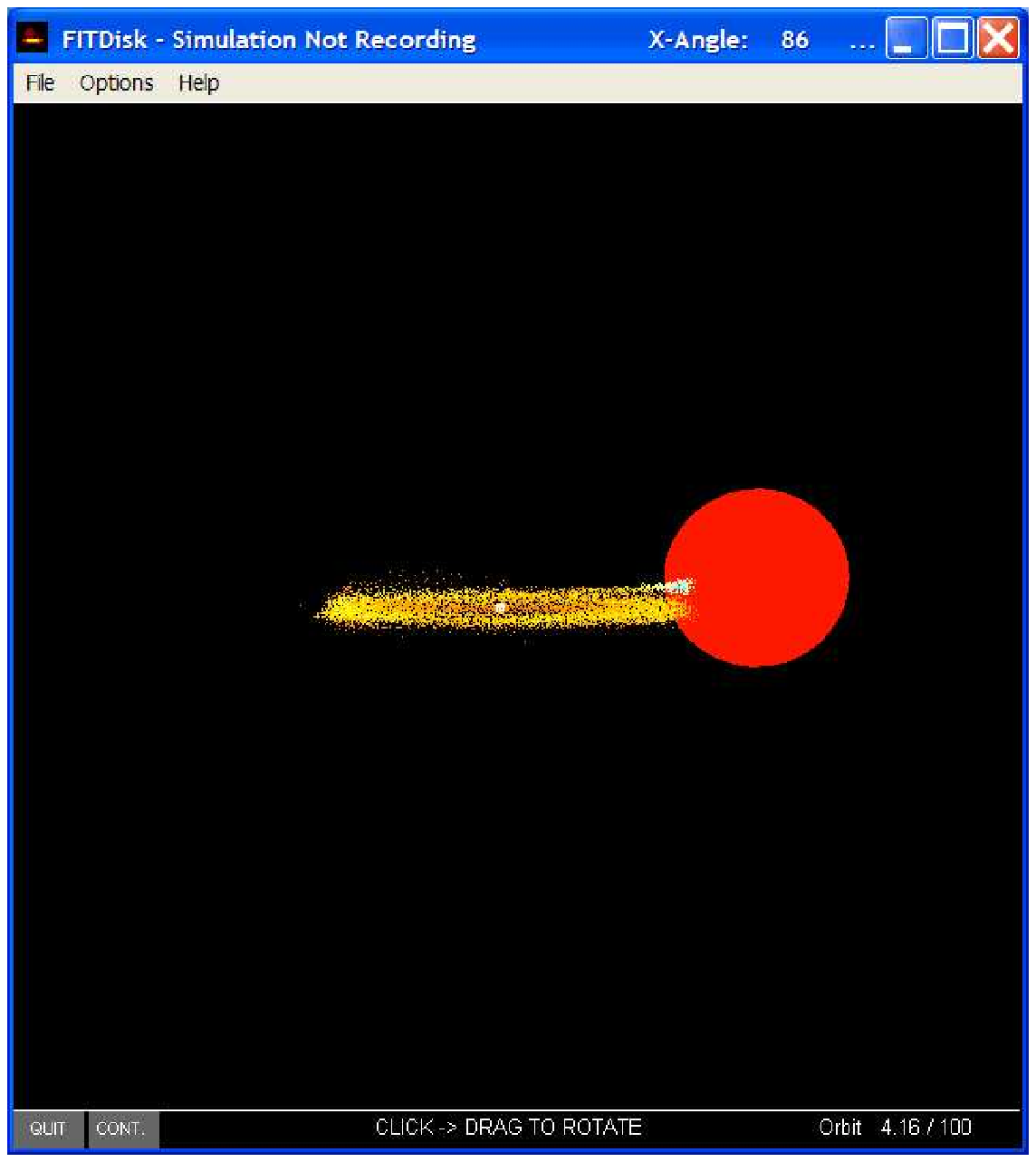}% Here is how to import EPS art
\caption{\label{fig:rotated} The simulation and renderings are fully three-dimensional.  At any time during the simulation, the user can pause the run, click-drag the mouse to change the inclination, and continue the run.  Note the rotation angle is indicated in the title bar.}
\end{figure}

Simulations are visualized with smooth, double-buffered graphics.  The view can be interactively rotated about a horizontal axis by clicking and dragging inside the window with either mouse button.  The simulation can be paused (see Figure\ref{fig:rotated}), allowing an opportunity to examine structural details or capture screenshots using the Windows PrntScrn key. To provide more insight into the structure and dynamics involved, the fluid particles are given false color based on their relative density or temperature as selected from the Options menu.  This feature gives an excellent qualitative picture of the radial density and temperature profiles, and also helps reveal features such as spiral shocks and superhumps.

\begin{figure}
\plotone{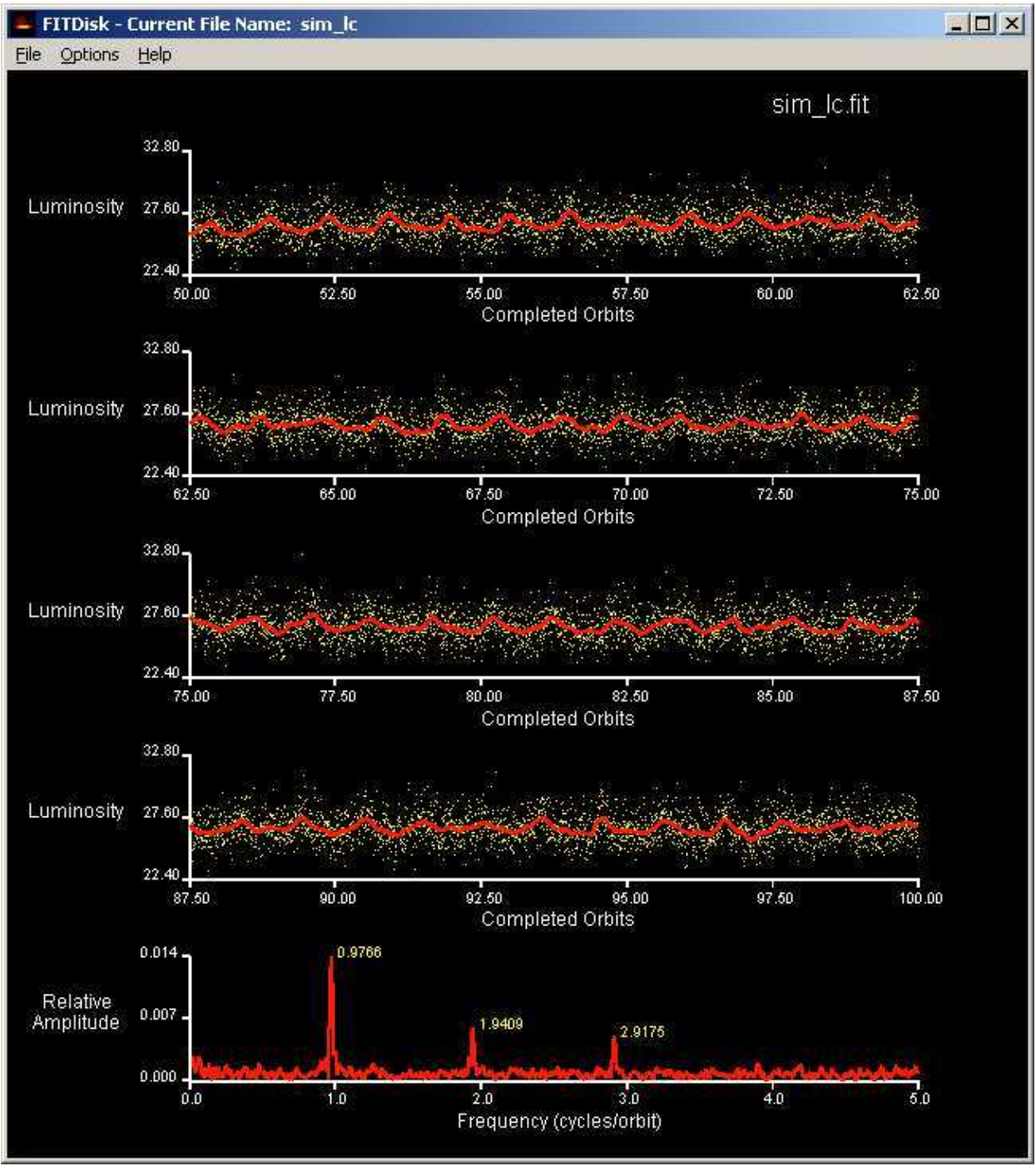}% Here is how to import EPS art
\caption{\label{fig:lc+fft} After a simulation has finished, the user can plot the simulation light curve and associated Fourier amplitude spectrum.  Peaks in the transform that are more than 5$\sigma$ above the mean are labeled.}
\end{figure}

Within the program, the frequencies present in the simulation light curves can be plotted and analyzed through use of an included fast Fourier transform (FFT) (see Figure\ref{fig:lc+fft}).  Time series plots show both the raw data and a boxcar-smoothed fit.  The smoothed curve can appear markedly similar to the light curves observed in real systems, as we discuss below.  The Fourier amplitude spectrum is only shown out to a frequency of five cycles per orbit to highlight the frequency range of interest.  Peaks in the FFT that are five or more standard deviations ($3\sigma$) from the mean are labeled with their specific frequencies.  In this way, for superhumping systems, the program often identifies the fundamental superhump frequency as well as the second and third harmonics.  Thus, users can explore the relationship between the superhump oscillation period and the orbital period as a function of binary system mass ratio.

\begin{table}
\caption{\label{tab:table1} \FITDisk\ User-Adjustable Parameters}
\begin{tabular}{ccccl}
\tableline
Param.& \omit\hss\rm Min\hss & \omit\hss\rm Max\hss & \omit\hss\rm Default\hss & Notes\\
\tableline\tableline
$M_1$ & 0.5 & 1.3 & 1.0 & Solar Masses\\
$q$   & 0.01 & 1.0 & 0.15 & $M_2/M_1$\\
\# Orbits & 1 & $\infty$\footnote{No set maximum.} & 10 \\
\# Particles & 1 & 25,000 & 1,000 \\
\# Ndisp & 1 & $\infty$  & 1 & Timestep(s) b/n display \\
\# Ninj  & 1 & 10  & 10 & \# Inj per timestep\\
$h$ & 0.008 & 0.040 & 0.010 & Smoothing Length\\
$\alpha$ & 0.01 & 2.00 & 1.00 & Viscosity parameter\\
$\beta$ & 0.01 & 2.00 & 1.00 & Viscosity parameter\\
Rec. Control & Off & On & Off & Record to disk?\\
Rec. Start & 1 & $\infty$ & 1 & Rec. Start Orbit\\
Rec. Resolution & Low & High & High & Rec. Resolution\\
Rec. Light Curve & Off & On & Off & Output light curve?\\
Light Curve Start & 1 & $\infty$ & 1 & Light Curve Start Orbit\\
Restart Control & Off & On & Off & Write restart files?\\
Restart Rate N & 1 & $\infty$ & 1 & Orbits b/n write \\
Star Style & Points & Filled & Filled & Star 2 visualization\\
Particle Size & Small & Large & Small & \\
Particle Color & Density & Temperature & Density & \\
\tableline
\end{tabular}
\end{table}

When File$\rightarrow$New Disk is selected, the simulations begin with no particles present.  The particles are injected in a burst at the $L_1$ point.  The stream initially travels on a roughly ballistic trajectory until the lead particles intersect the stream. Viscosity quickly causes the disk to collapse to the circularization radius, which has the same specific angular momentum as the particles entering at the $L_1$ point.  Over the next several orbits as the mass burst continues, viscosity acts to spread the mass out in both directions from the circularization radius.  If the mass ratio $q \le 0.34$ and the viscosity parameters $\alpha$ and $\beta$ are of order unity, then eventually superhump oscillations will be driven into resonance by the rotating tidal field of the secondary. 

If the restart option is checked, then at a rate determined by the user the code will write out a file that can be read in so that the user can continue the simulation from that point.  This might be useful if the user didn't specify sufficient orbits for superhumps to develop in a high-resolution simulation run. 

\section{Comparison with Observations}

\begin{figure}
\plotone{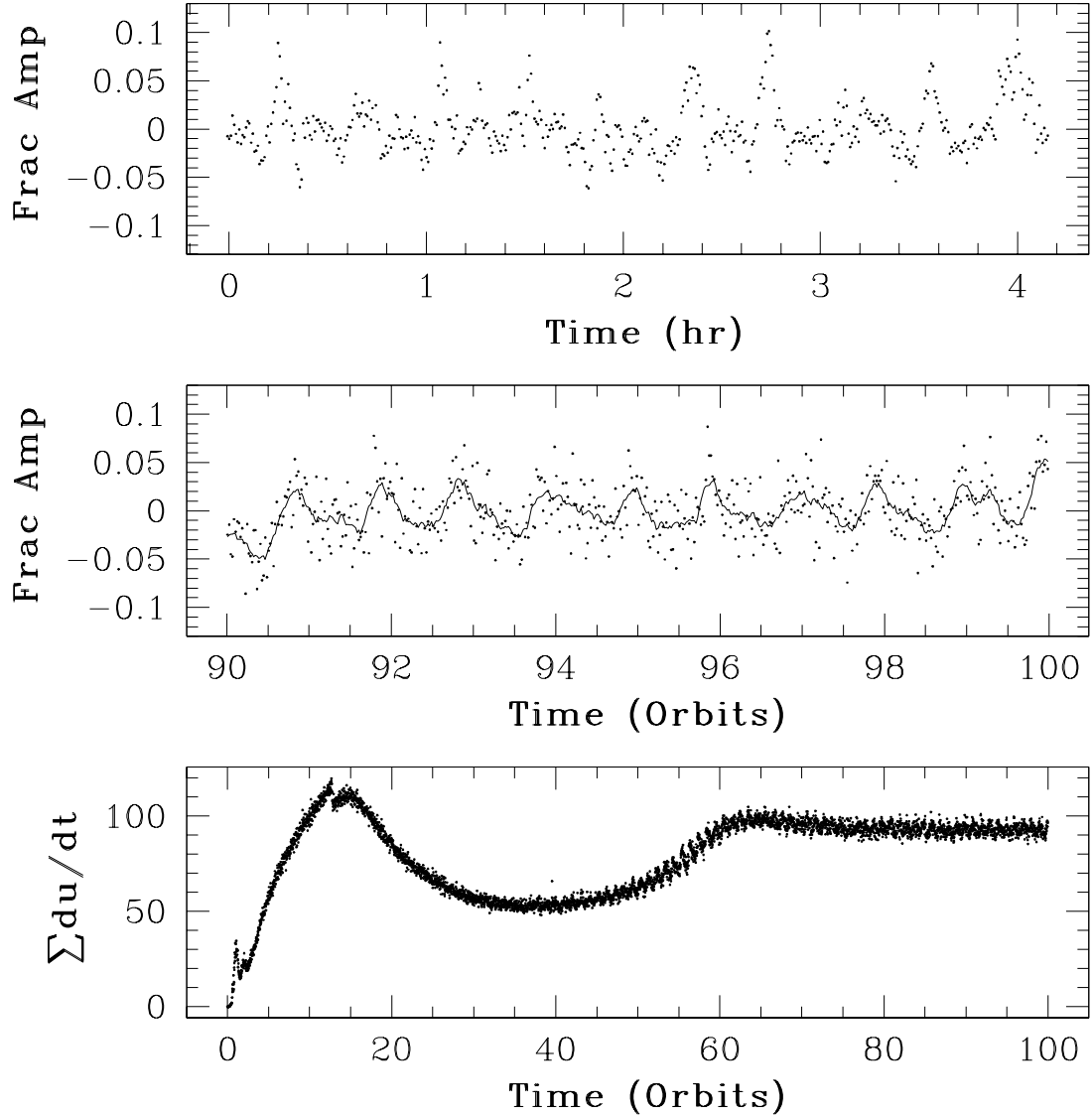}% Here is how to import EPS art
\caption{\label{fig: comp} A comparison of the light curve of the helium dwarf nova CR Boo with simulation results.  The top panel shows approximately 10 superhump cycles of CR Boo.  The middle panel shows the final 10 orbits of a $q=0.05$ simulation, where a boxcar-smoothed curve is included to guide the eye.  The bottom panel shows the entire simulation light curve.  }
\end{figure}

Isaac Silver (Florida Inst. Tech) observed the interacting binary white dwarf CR Boo (Wood et al. 1987, Patterson et al. 1997) with the 0.9-m SARA\footnote{Southeastern Association for Research in Astronomy, {\tt www.saraobservatory.org}.} telescope located at Kitt Peak National Observatory on the night of 2005 March 10 (UT) using an Apogee AP7P CCD camera with no filter to maximize photon counts.  The exposure time was 20 s, with an 11 s readout time.  The data were reduced using IRAF\footnote{IRAF is distributed by the National Optical
   Astronomy Observatories, which are operated by the Association of
   Universities for Research in Astronomy, Inc., under cooperative
   agreement with the National Science Foundation.}.  The top panel of Figure~\ref{fig: comp} shows $\sim$4.2 hr ($\sim10$ orbits) of time series data for CR Boo, which has an orbital period of $P_{\rm orb} = 1471$ s, and a superhump period of $P_{\rm sh} \approx 1490$ s (Patterson et al. 1997).  Note that the observed superhump signal varies considerably from cycle to cycle.
   
To demonstrate that \FITDisk\ generates useful output to compare with observations, we simulated a CV with mass ratio $q=0.05$ for 100 orbits using 25,000 particles, $h=0.02 a$, and viscosity parameters $\alpha=1.5$ and $\beta = 1.0$.  The bottom panel of the Figure shows the entire simulation light curve, and the middle panel shows the final 10 orbits.  Because the simulation is limited to 25,000 particles, the noise band is larger than the observations.  To compare with observations, we have first summed 4-points so the sample rate per orbit is approximately that of the top panel, and we then include a boxcar-smoothed line ($\rm width = 11$ points) to guide the eye.  Although we have made no special effort to match the observed light curve, the comparison of the simulation output with the observed light curve demonstrates the promise of using \FITDisk\ to constrain system parameters.

\section{Conclusion}

Cataclysmic variables are mass-transfer binary star systems that display a rich variety of behavior in their accretion disks.  As a class they have long been a favorite target of professional and amature astronomers alike.  Time-series photometry of their outbursts and superhump disk oscillations coupled with detailed numerical simulations have the potential to reveal not just the structure and evolution of the disks via seismology, but also perhaps the fundamental origin of viscosity in differentially rotating astrophysical plasmas.  

We present here the freely-available precompiled-binary code \FITDisk\ which allows non-specialist users (e.g., amature astronomers, students, and classroom instructors) to simulate the dynamics of cataclysmic variable accretion disk dynamics, including superhump oscillations, over a large volume of parameter space.  The code is a demonstration version of our fully three-dimensional smoothed particle hydrodynamics (SPH) code (Simpson \& Wood 1998; Wood, Montgomery \& Simpson 2000; Wood et al. 2005), with an intuitive, easy-to-use graphical user interface added on and ported to the dominant operating system for PCs.  It is our hope that this code will find wide use in not only amature and professional astronomical circles, but also in educational settings ranging from high schools to universities.

\begin{acknowledgments}
We are grateful to Isaac Silver for providing his unpublished time-series observations of CR Bootis.  This work was supported in part by grant AST-0205902 to The Florida Institute of Technology.
\end{acknowledgments}

\end{document}